\documentclass[graybox]{svmult}

\usepackage{mathptmx}       
\usepackage{helvet}         
\usepackage{courier}        
\usepackage{type1cm}        
\usepackage{makeidx}         
\usepackage{graphicx}        
\usepackage{multicol}        
\usepackage[bottom]{footmisc}

\usepackage{tikz}
\usepackage{paralist}

\usepackage{url}

\newcommand{\kinectTM}{Kinect\texttrademark}
\newcommand{\kinectTMS}{\kinectTM~}

\graphicspath{{./}}

\makeindex    
\begin{document}

\title*{Asymmetric pedestrian dynamics on a staircase landing from continuous measurements}
\author{Alessandro Corbetta,  Chung-min Lee,  Adrian Muntean and Federico Toschi}
\institute{Alessandro Corbetta \at Eindhoven University of Technology, 
  Eindhoven, The Netherlands and Politecnico di Torino, 
  Turin, Italy, \email{a.corbetta@tue.nl}
\and Chung-min Lee \at California  State University Long Beach, Long Beach, CA, USA
\and Adrian Muntean \at Karlstad University,
Karlstad, Sweden
\and Federico Toschi \at Eindhoven University of Technology,
Eindhoven, The Netherlands and  CNR-IAC,
Rome, Italy
}

\maketitle

\abstract*{We investigate via extensive experimental data the dynamics of pedestrians walking in a corridor-shaped landing in a building at Eindhoven University of Technology. With year-long automatic measurements employing a Microsoft \kinectTMS 3D-range sensor and \textit{ad hoc}  tracking techniques, we acquired few hundreds of thousands pedestrian trajectories in real-life conditions. Here we discuss the asymmetric features of the dynamics in the two walking directions with respect to the flights of stairs (i.e. ascending or descending). We provide a detailed analysis of position and speed fields for the cases of pedestrians walking alone undisturbed and for couple of pedestrians in counter-flow. Then, we show average walking velocities exploring all the observed combinations of numbers of observed pedestrians and walking directions.
}

\abstract{We investigate  via extensive experimental data the dynamics of pedestrians walking in a corridor-shaped landing in a building at Eindhoven University of Technology. With year-long automatic measurements employing a Microsoft \kinectTMS 3D-range sensor and \textit{ad hoc}  tracking techniques, we acquired few hundreds of thousands pedestrian trajectories in real-life conditions. Here we discuss the asymmetric features of the dynamics in the two walking directions with respect to the flights of stairs (i.e. ascending or descending). We provide a detailed analysis of position and speed fields for the cases of pedestrians walking alone undisturbed and for couple of pedestrians in counter-flow. Then, we show average walking velocities exploring all the observed combinations in terms of numbers of pedestrians and walking directions.}

\section{Introduction}
\label{sect:intro}
During the last two decades experimental investigations of pedestrians dynamics flourished, fostering a transition from qualitative to quantitative analyses. Several geometric configurations and flow scenarios have been studied in controlled laboratory conditions, such as corridors, bottlenecks, intersections and T-junctions dynamics~\cite{DBLP:journals/ijon/BoltesS13,zhang2011transitions,zhang2014comparison}. 
More recently, 3D-range cameras and wireless sensors enabled reliable  measurements in real-life conditions~\cite{Brscic201477,corbetta2014TRP,seer2014kinects,roggen2011recognition}, allowing for  data collection  with reduced (potential) influences of laboratory environments. Notably, these technologies are privacy-safe,  as recorded pedestrians are not identifiable, thus,  unlimited data collections, e.g., via long term measurement campaigns~\cite{corbetta2014TRP} are possible. 

In this paper we analyze the dynamics of pedestrians in a landing  (intermediate planar area between flights of stairs) which has corridor-like geometry. Few experimental data have been collected in these scenarios, typically in the context of evacuation dynamics~\cite{hoskins2012differences,ronchi2014analysis}.
Driven by fundamental curiosity, we recorded the landing on a 24/7 basis and acquired the trajectories of walking pedestrians in a year-long experimental campaign.  Our data include multiple natural traffic scenarios such as uni- or bi-directional flows with one or several pedestrians. After categorizing the measurements based on walking directions and number of pedestrians involved, we compare pedestrian positions and velocities among different flow conditions. 
We note that individuals walking in a landing are either ascending or descending the neighboring stair flights, and this aspect induces asymmetries in the dynamics, likely related to the different physical fatigue of pedestrians. These asymmetries that we observe here and discuss appear on side of cultural preferences, for instance for the  walking side~\cite{moussaid2009experimental}.

This paper is organized as follows: in Sect.~\ref{sect:Meas} we provide a description of our measurement setup and a primer of the recording technique. In Sect.~\ref{sect:ped} we give a detailed overview of the dynamics of pedestrians walking alone and in avoidance of one other individual via  position and velocity fields. Moreover, we comment on the average velocities considering all the possible flow conditions and addressing all  direction combinations. A concluding  discussion is reported in Sect.~\ref{sect:concl}.

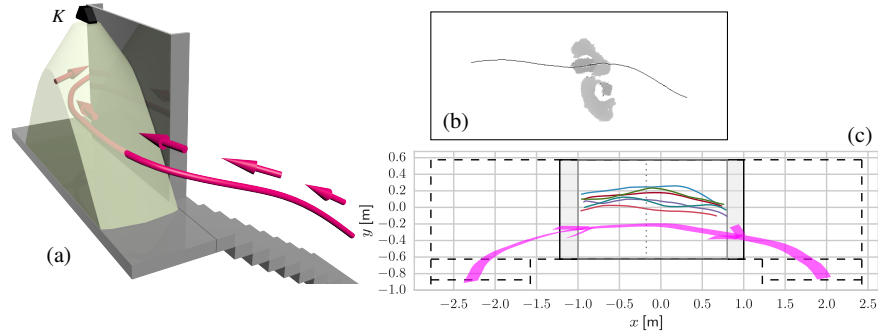
\begin{figure}[t]
\begin{tikzpicture}

\node at (0.\textwidth,-1.) {\includegraphics[width=.4\textwidth,trim=10.5cm 4.95cm 4.3cm 4.3cm, clip=true]{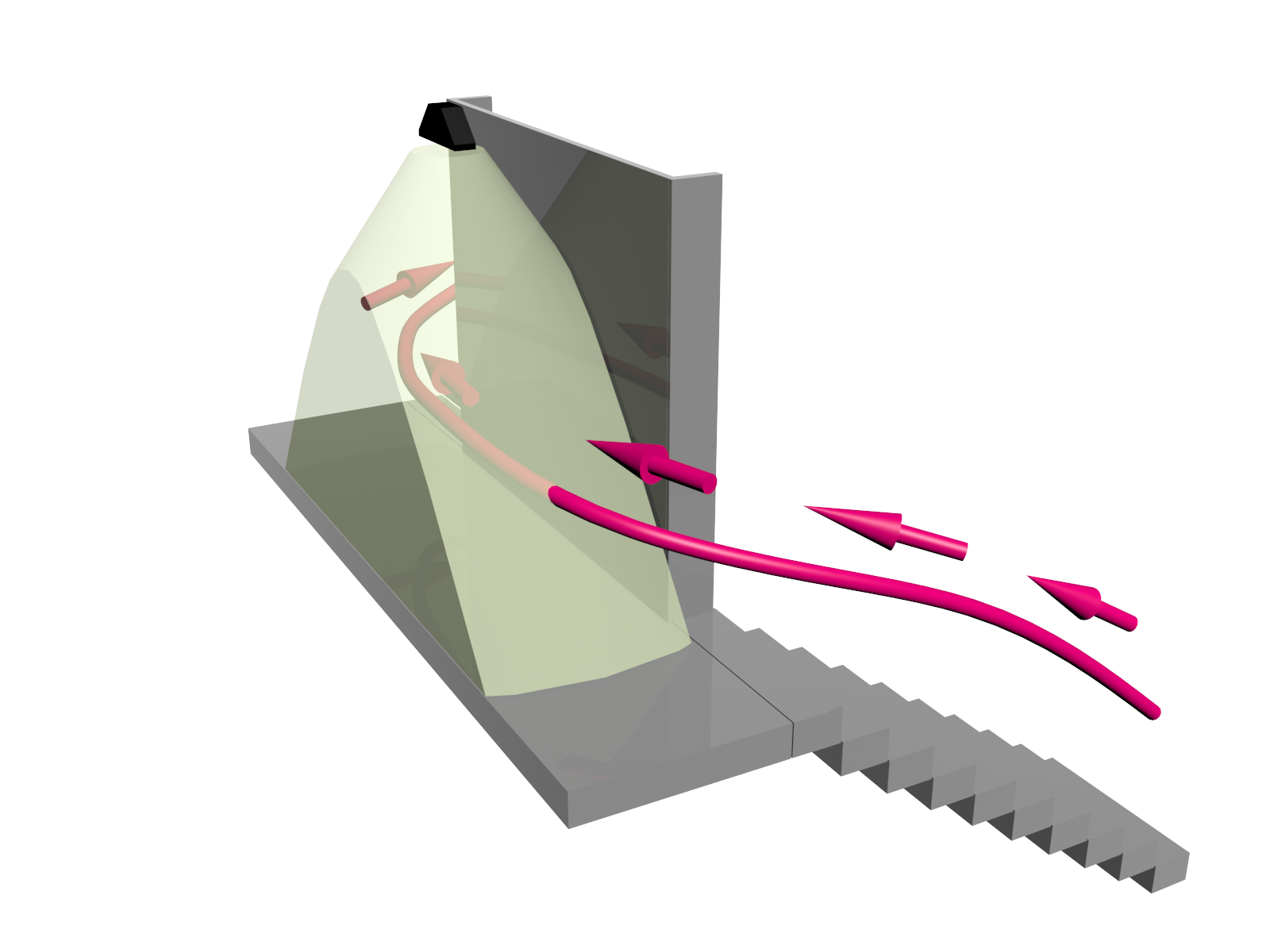}};
\node at (-1.65,.7) {\scriptsize$K$};
\node at (.5\textwidth,-2.2) {\includegraphics[width=.6\textwidth,trim=.5cm 2.4cm 1.3cm 2.8cm, clip=true]{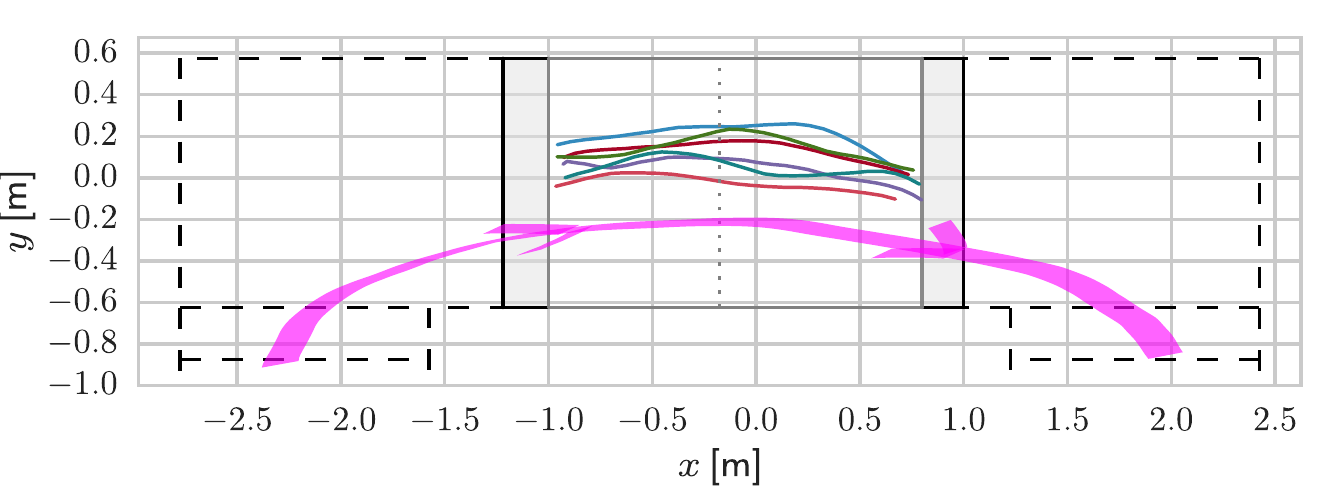}};
\node at (.45\textwidth,-.1) [draw=black] {\includegraphics[width=.32\textwidth,trim=1.4cm 5.cm 0.9cm 3.8cm,clip=true]{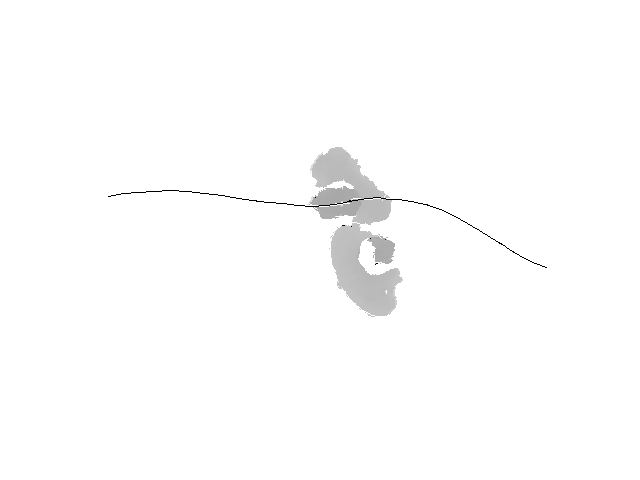}};
\node[fill=white] at (-1.65,-2.5) {(a)};
\node at (.31\textwidth,-.725) {(b)};
\node at (9.,-.9) {(c)};
\end{tikzpicture}
\caption{(a) Sketch of the landing including the view cone of the \kinectTMS sensor (marked with a letter ``K''). (b) A frame taken in the landing by the \kinectTMS sensor; two  pedestrians walking in opposite directions are present. The depth field ($z$) is represented via the gray scale. Brighter pixels are farther from the camera plane. (c)  Planar view of the landing with dimensions and the $xy$ reference system considered. The walking direction from the zeroth to the first floor (from left to right) is depicted. Examples of trajectories collected are reported.   
}
\label{fig:experiment-general}
\end{figure}

\section{Measurement site}\label{sect:Meas}
We measured the pedestrian traffic in a landing within the Metaforum building at Eindhoven University of Technology. The landing connects the two staircases in the configuration presented in  Fig.~\ref{fig:experiment-general}a and c, where individuals ascend in a clockwise direction from the zeroth 
to the first floor  of the building. The landing is $5.2\,$m long and $1.2\,$m wide, and the steps have the same width. Individuals at the zeroth floor reach the landing after $18$ steps, then they climb $4$ further steps arriving the first floor.
Pedestrian traffic mainly comes from students walking between the canteen of the building (zeroth floor) to the dining area (first floor) and \textit{vice versa}. Considering the reference system in Fig.~\ref{fig:experiment-general}c, we  indicate the walking direction that leads to the first floor as \textit{left to right} (2R, for brevity) and as \textit{right to left} (2L) the opposite case.  On average, $2.200\,$ pedestrians cross the facility every working day,  and occupancy peaks at around 12\,PM  (lunch time) and at around  4\,PM (afternoon break). At peak hours, typically there are multiple pedestrians walking in the facility  (up to six pedestrians have been recorded in our observation window at once) in co-flow (uni-directional dynamics) or counter-flow (bi-directional dynamics). Conversely, off-peak traffic is mostly due to individuals walking alone, \textit{undisturbed} by other pedestrians.  We refer to our previous work~\cite{corbetta2014TRP} for time histories and statistics about daily traffic.
In this work, we discuss pedestrian dynamics data acquired during 107 working days in the period October 2013 -- October 2014. In this campaign we collected \textit{ca.} 230.000 time-resolved high-resolution trajectories. 

\bigskip

\noindent\textbf{Data acquisition.} We measured  trajectories of pedestrians via an automatic head tracking procedure that allows non-intrusive and privacy respecting data acquisition in real-life condition.  Such procedure is based on  the 3D-depth data delivered by an overhead and downward looking Microsoft \kinectTMS 3D-range sensor. 3D-depth frames represent a filmed scene as a three dimensional $(x,y,z)$  pixel \textit{cloud} (cf. Fig.~\ref{fig:experiment-general}b).
Pedestrians identification (segmentation) can be operated by identifying and isolating pixel \textit{clusters} within such a cloud. Heads, that we track as particles, are  the topmost portions of each cluster (cf.~\cite{brscic2013person,seer2014kinects}). 

We filmed at $15$ frames per second in the central, $1.8\,$m long (cf. Fig.~\ref{fig:experiment-general}c), section of the landing, by placing a \kinectTMS sensor at an height of \textit{ca.} $4\,$m (cf. Fig.~\ref{fig:experiment-general}a). Technical aspects of our detection approach, inspired by~\cite{seer2014kinects}, are discussed in the appendix of~\cite{corbetta2015MBE}. Furthermore, we employed the OpenPTV library~\cite{OpenPTV}, developed by  the Particle Tracking Velocimetry~\cite{willneff2003spatio}  community in fluid mechanics, to perform heads tracking and to retrieve trajectories (cf. Fig.~\ref{fig:experiment-general}c).

\section{Pedestrian dynamics}\label{sect:ped}
The U-shape of the landing influences the dynamics of pedestrians that follow curved trajectories to reach the staircase at the opposite end of the walkway. Hence, contrarily from what is expected in a rectilinear corridor of similar size, pedestrian positions and velocities are asymmetric in space. These aspects depend on the flow conditions (undisturbed pedestrian vs. multiple pedestrians) as well as on the walking directions (ascending vs. descending).

\begin{figure}[t]
\sidecaption[t]
\begin{tikzpicture}
\node at (0,0) {\includegraphics[width=.62\textwidth,trim=.3cm 1.6cm 1.5cm 3.1cm, clip=true]{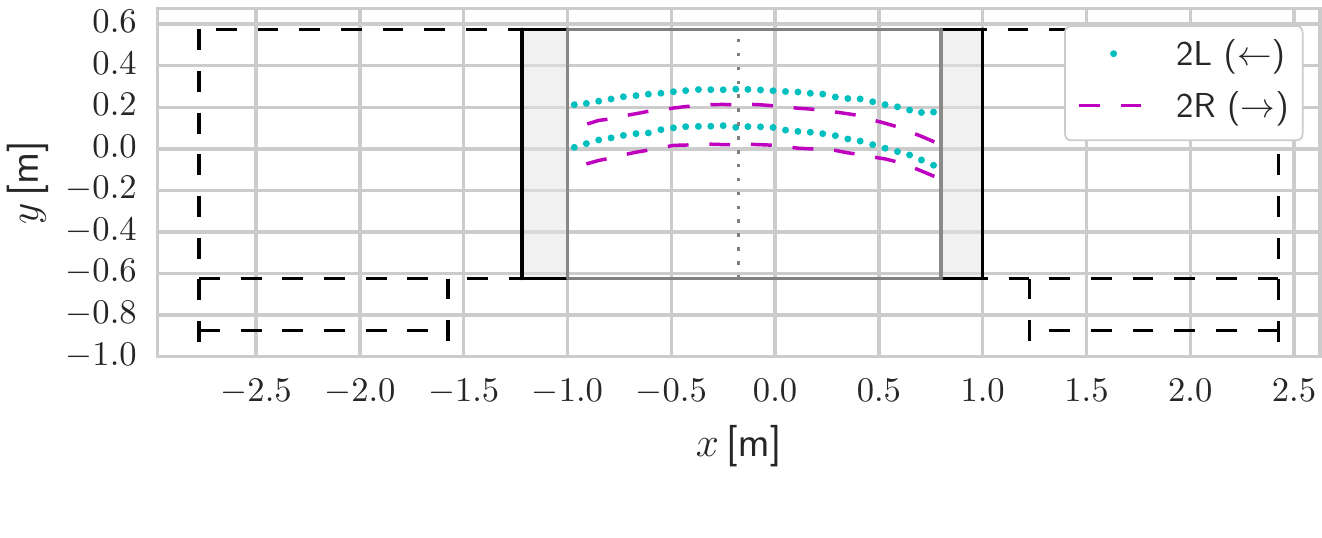}};
\node[fill=white] at (-3,-.8) {(a)};
\node at (0,-3.8) {\includegraphics[width=.62\textwidth,trim=1.82cm 1.5cm 2.8cm .9cm, clip=true]{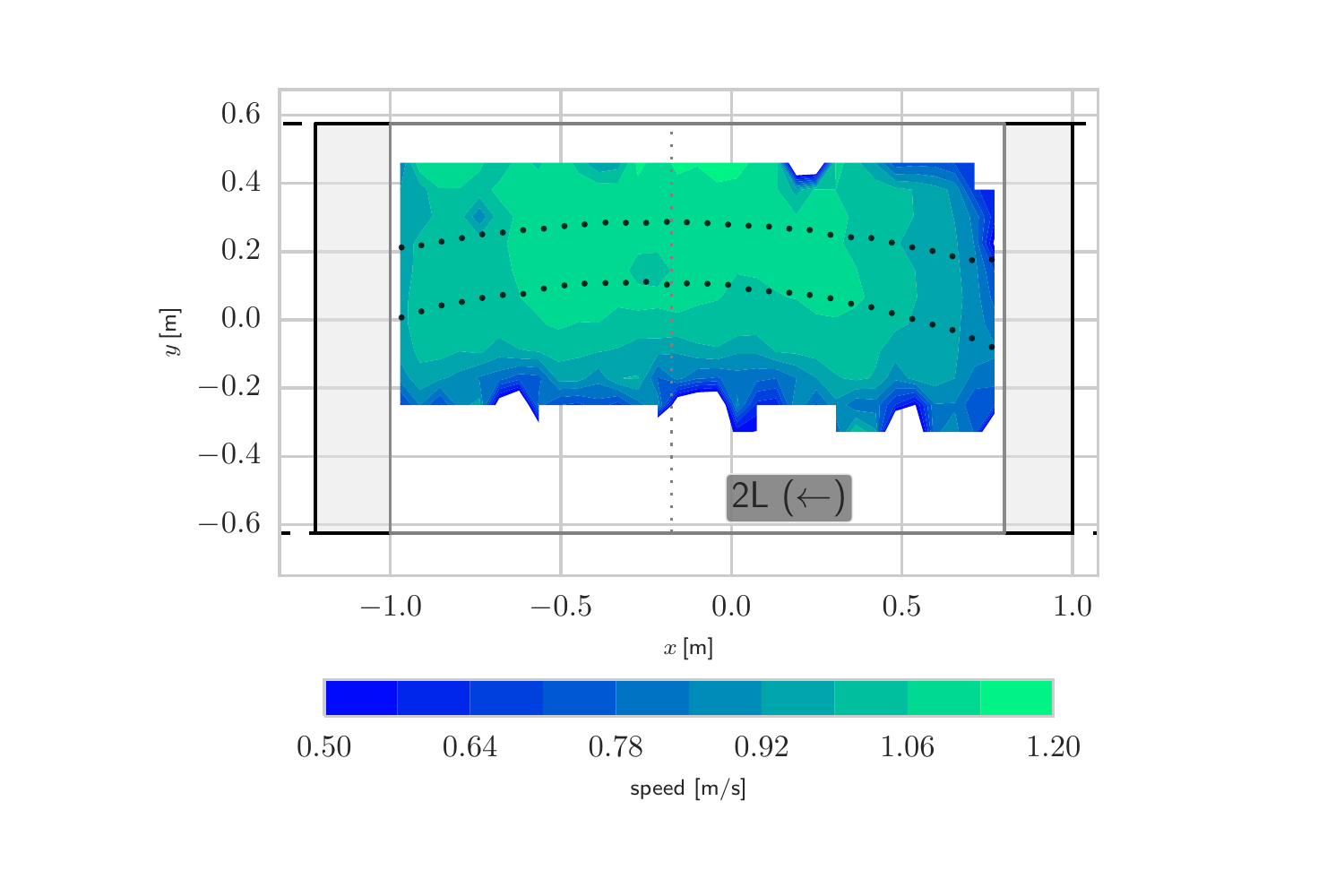}};
\node at (0,-7.9) {\includegraphics[width=.62\textwidth,trim=1.82cm 1.cm 2.8cm 1.cm, clip=true]{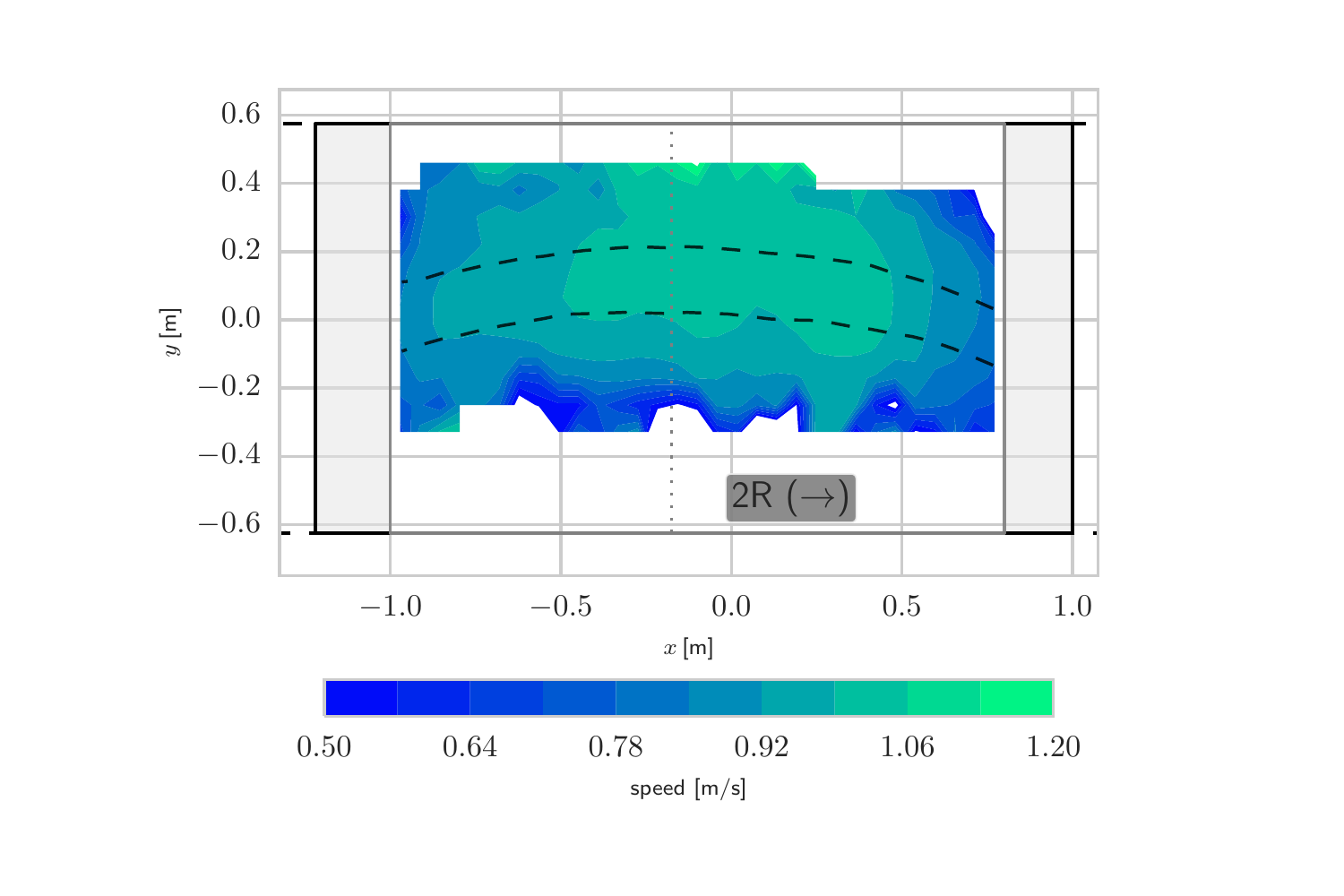}};
\node[fill=white] at (-3,-4.95) {(b)};
\node[fill=white] at (-3,-9.) {(c)};
\end{tikzpicture}
\caption{Positions and velocities of  pedestrians walking undisturbed. (a) Positions concentrate mostly in thin curved layers following the U-shaped geometry. To evaluate these layers we address separately pedestrians going from left to right (2R, for brevity) and from right to left (2L). For each ``horizontal'' location $x$ in the observation window ($x\in [-1,0.8]\,$m), we consider the distribution $y_x$ of pedestrian positions in  ``vertical'' direction. We report the $15^{th}$ and the $85^{th}$ percentiles of $y_x$ as a function of $x$ (thus the vertical interval $[y_{x,15},y_{x,85}]$).  Layers for pedestrians going to the left and to the right are identical but a $\approx 20\,$cm vertical offset. (b,c) Fields of average walking speed in space. Respectively for pedestrians going to the left (b) and to the right (c). In both cases the maximum velocity (higher for pedestrians going to the left, that have already descended a ramp of stairs) are reached after the central part of the corridor. Thus, pedestrians  decelerate to approach the next ramp.
}
\label{fig:single-ped-landing}
\end{figure}
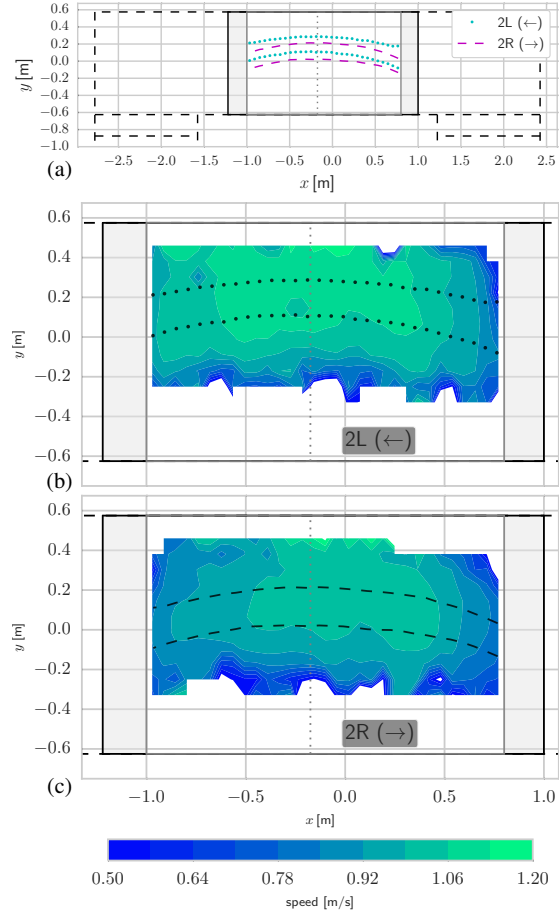

\begin{figure}[h]
\sidecaption[t]
\begin{tikzpicture}
\def\vtr{-.5};
\node at (0,0.) {\includegraphics[width=.62\textwidth,trim=.2cm 1.cm 1.5cm 3.cm, clip=true]{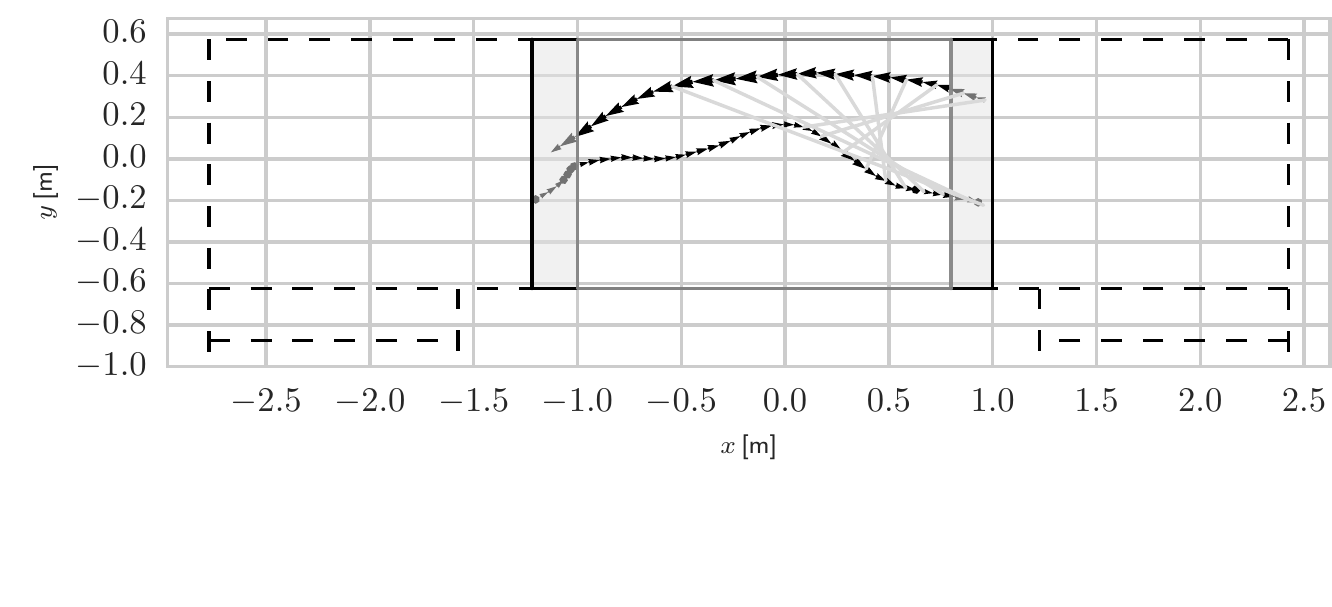}};
\fill [white] (-3.5,-.4) rectangle (3.75,-1.5);
\node at (.0275,-2.4-\vtr) {\includegraphics[width=.6155\textwidth,trim=.3cm 1.6cm 1.5cm 3.1cm, clip=true]{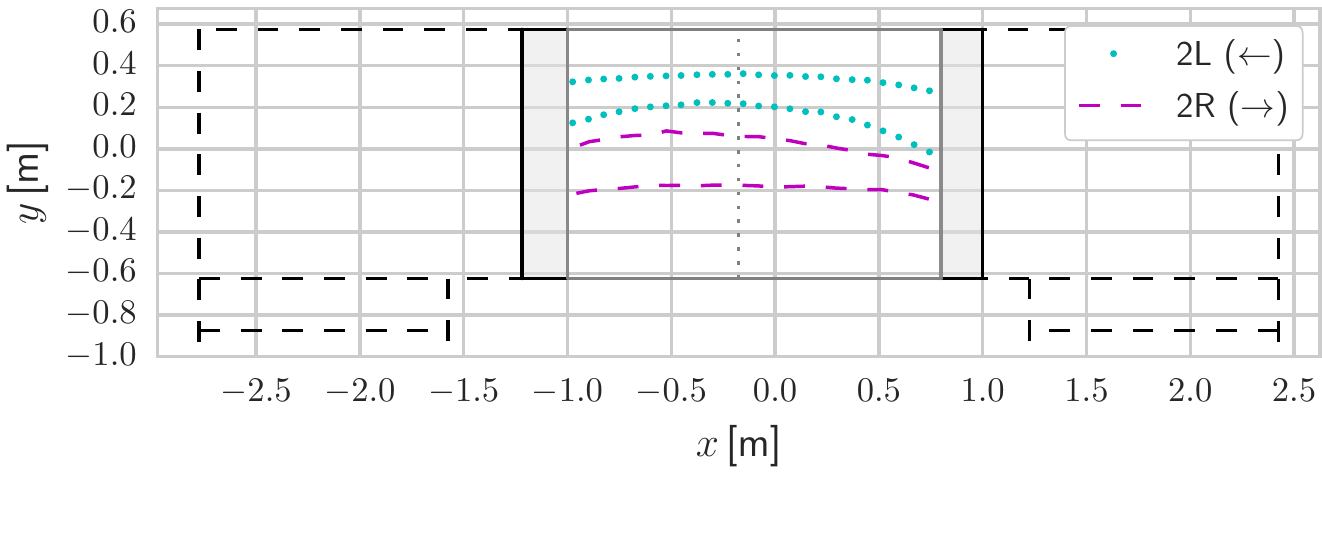}};
\node at (0,-6.4-\vtr) {\includegraphics[width=.62\textwidth,trim=1.82cm 1.cm 2.8cm 1.cm, clip=true]{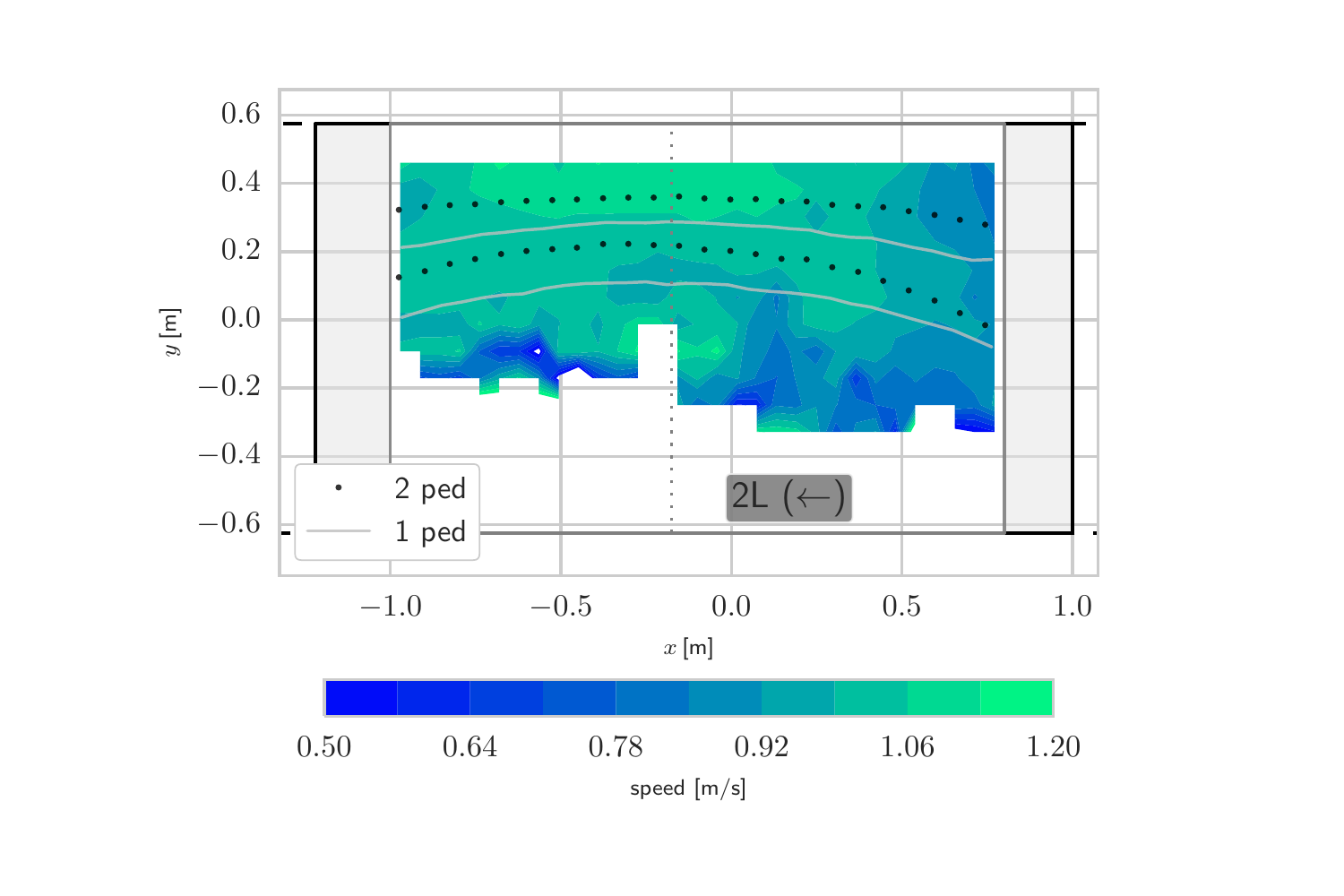}};
\node at (0,-10.25-\vtr) {\includegraphics[width=.62\textwidth,trim=1.82cm 1.cm 2.8cm 1.cm, clip=true]{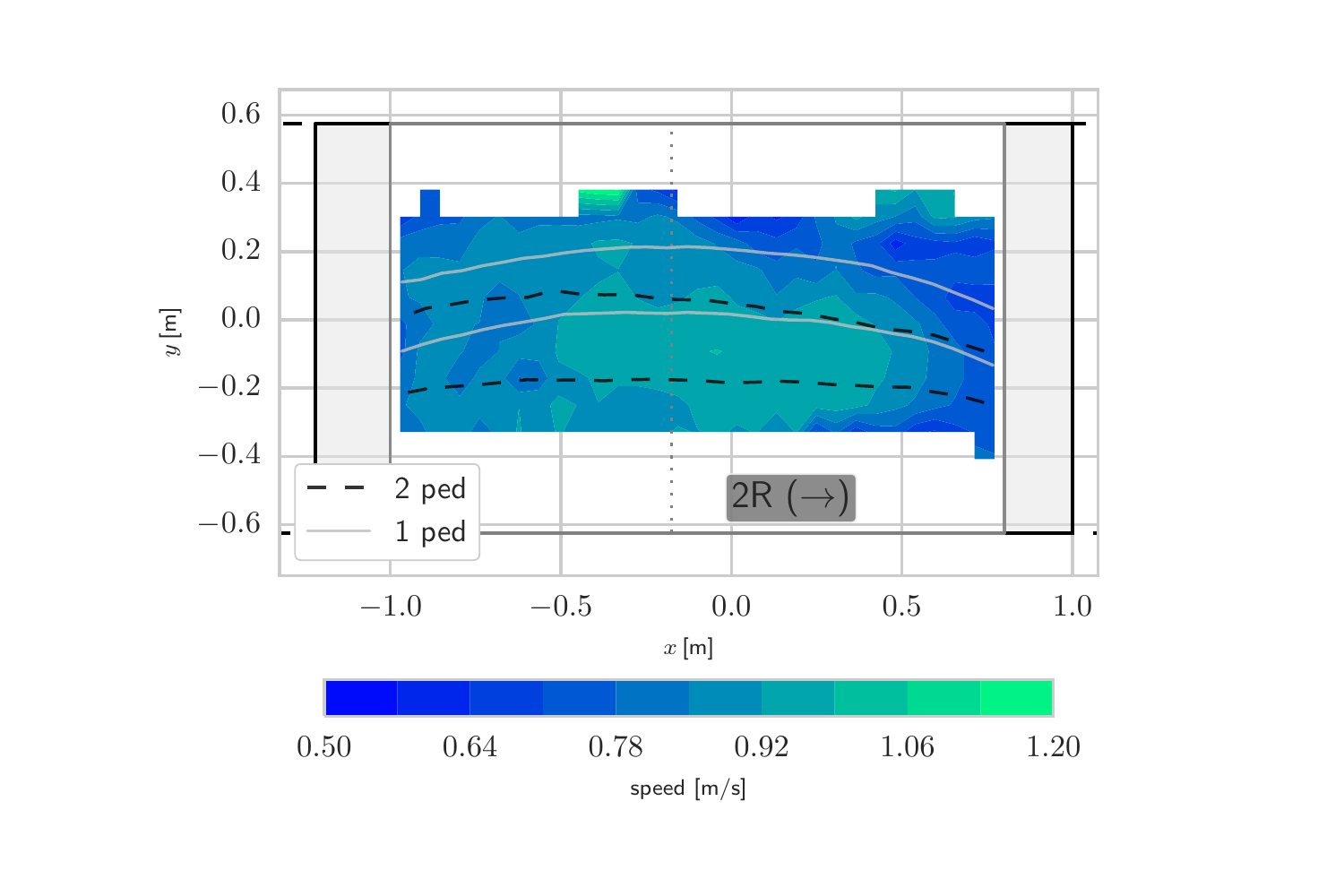}};
\node[fill=white] at (-3,-.65) {(a)};
\node[fill=white] at (-3,-3.2-\vtr) {(b)};
\node at (-3,-7.3-\vtr) {(c)};
\node[fill=white] at (-3,-11.4-\vtr) {(d)};
\end{tikzpicture}
\caption{Positions and velocities of pedestrians walking in presence of a peer having opposite velocity (counter-flowing). (a) Examples of trajectories in counter-flow. Simultaneous detections are connected via gray segments. The pedestrian going to the right enters first. When the pedestrian going to the left appears, he or she modifies the trajectories moving to the relative right for avoidance. (b) In avoidance regime, pedestrians positions concentrate on the relative right. The layer of preferred positions is calculated as in Fig.~\ref{fig:single-ped-landing}a. Notably, the symmetry with respect to the corridor ``vertical'' axis ($x\approx -0.1\,$m. Cf. Dotted gray line) is lost. (c,d) Average speed fields for pedestrians going from right to left (c) and from left to right (d) in presence of a second pedestrians going in opposite direction. The preferred positions layer is reported and compared with the preferred positions layer in case of  undisturbed pedestrians (cf. Fig.~\ref{fig:single-ped-landing}). Pedestrians going to the left and to the right have smaller walking speed than in the undisturbed case. Notably, pedestrians going to the right walk significantly more slowly.} 
\label{fig:2counterflow}
\end{figure}

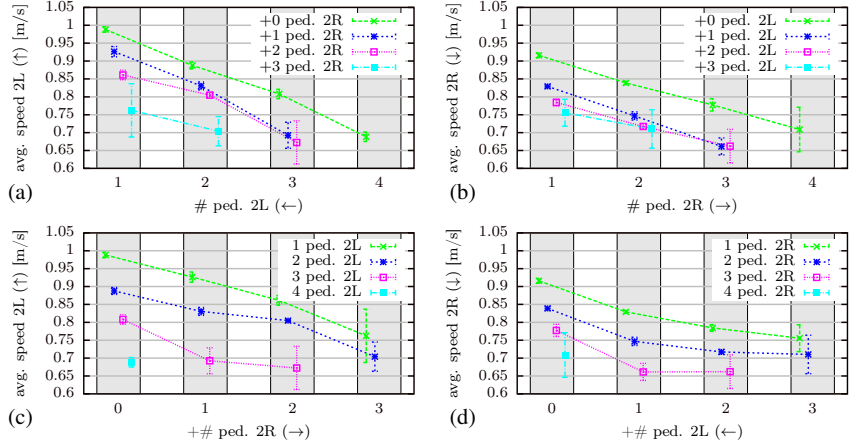
\begin{figure}[t]
\begin{tikzpicture}
\node at (0,0) {\includegraphics[width=.985\textwidth]{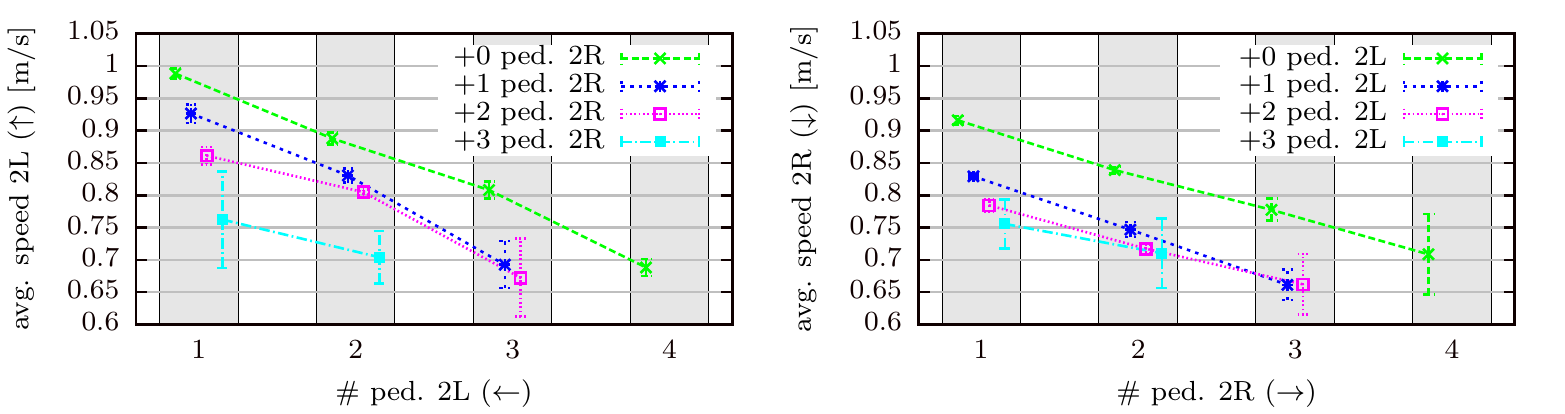}};
\node at (0,-3.) {\includegraphics[width=.985\textwidth]{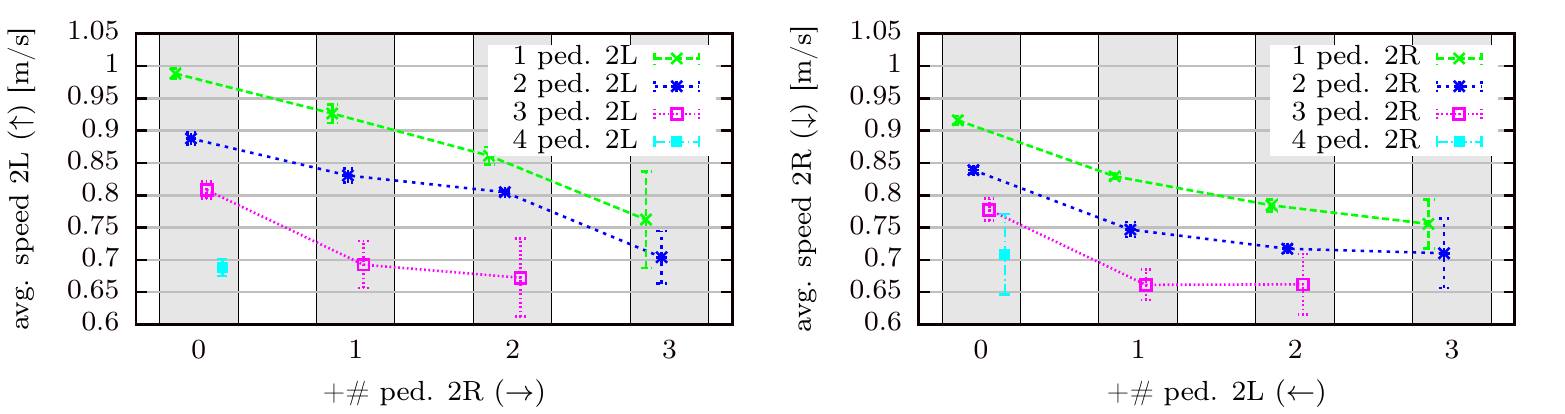}};
\node[fill=white] at (-5.6,-1.2) {(a)};
\node[fill=white] at (.28,-1.2) {(b)};
\node[fill=white] at (-5.6,-4.2) {(c)};
\node[fill=white] at (.28,-4.2) {(d)};
\end{tikzpicture}
\caption{Fundamental diagrams of different system state variable pairs (cf. Eq.~(\ref{eq:sys-state})). We report on the vertical axis average walking speeds for pedestrians going to the left (descending)  in (a,c) and for pedestrians going to the right (ascending) in (b,d). Average speeds are considered in dependence on the number of pedestrians in co-flow and counter-flow: in (a,b), the number of co-flowing pedestrians is on the horizontal axis, while the number of counter-flowing pedestrians is accounted via the different curves (for example, in (a) ``+2 ped. 2R'' means that there are two pedestrians going to the right in addition to a number of pedestrians going left). Diagrams (c,d) contain a ``transposed'' information, as the number of counter-flowing pedestrians is on the horizontal axis while the number of co-flowing pedestrians changes across the curves. We consider just system states for which there are at least $100$ frames. The size of the error bars (possibly underestimated) is $\max(s) - \min(s)$, where $s = \{s_1,s_2,s_3,s_4\}$, and the $s_i$ are average  values computed on a random even partition in four sets  of the speed data at a given (\# ped. 2L, \# ped. 2R) state. 
}
\label{fig:fdiagr-str}
\end{figure}

Direction-dependent differences can be observed already for pedestrians walking undisturbed. Pedestrian trajectories concentrate within thin curved layers that are located at the \textit{relative} right hand side of the facility (cf. Fig.~\ref{fig:single-ped-landing}a, the relative right hand side is at the absolute top for people going to the left and at the absolute bottom for people going to the right in the figure reference.  See the caption and cf.~\cite{corbetta2014TRP} for the layer estimation idea). These layers reflect a preferred walking path, ideally located along their axes, that acts as ``guiding center'' of trajectories fluctuations (cf.~\cite{CorbettaStochdyn} for analysis and modeling of such stochastic fluctuations). Although the relative position of the layers conforms with the cultural habit of keeping the driving side (cf., e.g.,~\cite{moussaid2009experimental}), an influence of the landing geometry cannot be excluded. In fact the shape of the landing limits the sight on the staircases, hence right hand side positions  may be kept to ease potential collisions (cf. Fig.~\ref{fig:2counterflow}). Walking speed is affected by the walking direction too: pedestrians descending from the stairs walk faster (cf. quantitative comparisons in Fig.~\ref{fig:fdiagr-str}
).  The walking speed varies in space and its contours are roughly transversal with respect  to the position layers. The speed peaks around the central section of the corridor, and remains high in the second half of the walkway. Individuals walk slower  near the staircases to adapt their velocity to the  ascent/descent of the stairs (a speed drop of about  $30\%$ is measured in our observation window, cf. Fig.~\ref{fig:single-ped-landing}b and c).

Direction-dependent differences increase when the presence other pedestrians trigger avoidance mechanisms. The simplest avoidance scenario involves exactly two pedestrians walking in opposite direction (i.e., counter-flowing. Cf. Fig.~\ref{fig:2counterflow}a). In this condition, the  path layers are shifted to the relative right to avoid collision. Contrary to the single pedestrian case, these layers have no overlap (cf. Fig.~\ref{fig:2counterflow}b vs. Fig.~\ref{fig:single-ped-landing}a), furthermore they are not symmetric with respect to the central corridor  vertical axis ($x\approx -0.1\,$m). In both 2L and 2R cases, layers are wider near the entrance side with similar distribution to  the undisturbed pedestrian case. Moving across the landing the layers constrict and 
shift toward the relative right hand side.
 We observe a  drop in the walking speed in comparison with the undisturbed pedestrians, especially around the central horizontal axis ($y\approx 0\,$m) where collisions may potentially occur. Higher walking speed are reached at the
 relative right hand side of the pedestrians, 
 where collision are mostly  avoided.  Comparing the  counter-flow dynamics in pedestrian pairs with the undisturbed dynamics, we observe  further direction-related asymmetries:
\begin{inparaenum}[(i)]
\item positions shift to the relative right from the undisturbed case is larger for pedestrians ascending (2R);
\item the speed drop in counter-flow is larger for pedestrians ascending (2R).
\end{inparaenum}

When more than two pedestrians are present different  walking configurations are possible. Moreover, pedestrians may have social interactions (e.g. conversations) and move in groups (cf., e.g.,~\cite{PhysRevE.89.012811}), that may influence the walking behavior (we ignore here such possible influences). 
  We address the walking dynamics considering  average walking speeds in all possible uni- and bi-directional flow configurations. 
We refrain from spatial analyses and we take here velocity averages over the observation window (cf. Fig.~\ref{fig:single-ped-landing}b,c and Fig.~\ref{fig:2counterflow}c,d).  
We identify configurations considering the number of pedestrians going to the left (\textit{\# ped. 2L}) and the number of pedestrians going to the right (\textit{\# ped. 2R}).  After grouping  pedestrians that in each frame walk in the same direction, we evaluate their average speed (respectively, \textit{avg. speed 2L} and \textit{avg. speed 2R}). In other words, we give a simplified description of the system state through a tuple:
\begin{equation}\label{eq:sys-state}
(\mbox{\# ped. 2L},\  \mbox{\# ped. 2R},\ \mbox{avg. speed 2L},\ \mbox{avg. speed 2R}).
\end{equation}
Considering average speed vs. the number of pedestrians yields fundamental diagram plots, that we report  in Fig.~\ref{fig:fdiagr-str}.
We observe a twofold monotonic behavior (within error bar) with  directional dependence. First, the average speed of pedestrians decreases as  the number of pedestrians increases either in co-flow or in counter-flow situations. Second, average speeds of ascending pedestrians are lower than those of descending pedestrians for any given combination of co-flowing and counter-flowing pedestrians. 
However, while an increase of co-flowing pedestrians (for fixed number of counter-flowing individuals) yields nearly linear reductions of the average speed (cf. Fig.~\ref{fig:fdiagr-str}a,b), the trend for increasing the number of pedestrians in counter-flow is not linear (cf. Fig.~\ref{fig:fdiagr-str}c,d). We observe the following features:
\begin{inparaenum}[(i)]
\item the velocity response to the number of counter-flowing pedestrians is different in the cases of individuals going to the left and going to the right, and
\item specifically for the population going to the right, significant speed drops occur as soon as one counter-flowing pedestrian is present; the exact number of counter-flowing individuals seems instead to play a minor role.
\end{inparaenum}

\section{Discussion}\label{sect:concl}
We acquired experimentally and in real-life conditions a large set of  trajectories of pedestrians walking in a landing. The trajectories span over multiple flow conditions involving a variable number of pedestrians walking in different direction configurations; in particular, both co-flows and counter-flows occur and are recorded. The U-shape of the landing as well as the previous ascent/descent of the stairs induce asymmetries in the dynamics that add up with cultural walking side preferences.  Pedestrians walking undisturbed keep the relative right side, even if no avoidance is necessary. This cultural preference is likely enhanced because of the limited vision near the staircases, which yields a choice of positions  preventing possible inbound collisions.
We considered average walking speed for all possible combinations of occupancy and walking directions. Pedestrians that have climbed the stair case (going to the right) appear to move slower than those who just descended for all flow configurations. Interestingly, the increment of co-flowing pedestrians yields nearly linear speed reductions, while this is not true when the number of counter-flowing pedestrians increases.

\begin{acknowledgement}
We  thank A. Holten and G. Oerlemans (Eindhoven, NL) for their help in the establishment of the measurement setup at Eindhoven University of Technology and  A. Liberzon (Tel Aviv, IL) for his help in the adaptation of the OpenPTV library. We acknowledge Iker Zuriguel (Pamplona, SPA) for the discussions during the TGF '15 conference that led to Fig.~\ref{fig:fdiagr-str}.  We acknowledge the support from the Brilliant Streets research program of the Intelligent Lighting Institute at the Eindhoven University of Technology, NL. AC was founded by a Lagrange Ph.D. scholarship granted by the CRT Foundation, Turin, IT and by Eindhoven University of Technology, NL.
\end{acknowledgement}

\bibliographystyle{spmpsci.bst}
\bibliography{bibliog}

\end{document}